\documentclass[twocolumn,english,aps,arxiv,superscriptaddress]{revtex4-1}
\usepackage{grffile}
\usepackage[latin9]{inputenc}
\usepackage{color}
\usepackage{babel}
\usepackage{graphicx}
\usepackage{epstopdf}
\usepackage{cancel}
\usepackage{amssymb}
\PassOptionsToPackage{normalem}{ulem}
\usepackage{ulem}
\setcitestyle{open={(},close={)}}
\usepackage[unicode=true, bookmarks=false, breaklinks=false,pdfborder={0 0 1},colorlinks=false] {hyperref}
\usepackage{breakurl}
\begin{document}

%\title{Unconventional magnetic phase diagram in a family of van der Waals magnets}
\title{Unconventional pressure-driven metamagnetic transitions in topological van der Waals magnets}
%\title{Independent probing of the three main magnetic interactions in MnBi$_4$Te$_7$}

\author{Tiema Qian$^\dagger$}
\affiliation{Department of Physics and Astronomy and California NanoSystems Institute, University of California, Los Angeles, Los Angeles, CA 90095, USA}
\author{Eve Emmanouilidou$^\dagger$}
\affiliation{Department of Physics and Astronomy and California NanoSystems Institute, University of California, Los Angeles, Los Angeles, CA 90095, USA}

\author{Chaowei Hu}
\affiliation{Department of Physics and Astronomy and California NanoSystems Institute, University of California, Los Angeles, Los Angeles, CA 90095, USA}

\author{Jazmine C. Green}
\affiliation{Department of Physics and Astronomy and California NanoSystems Institute, University of California, Los Angeles, Los Angeles, CA 90095, USA}

\author{Igor I. Mazin}
\affiliation{Department of Physics and Astronomy, George Mason University, Fairfax, VA 22030, USA}
\affiliation{Quantum Science and Engineering Center, George Mason University, Fairfax, VA 22030, USA}

\author{Ni Ni}
\email{Corresponding author: nini@physics.ucla.edu}
\affiliation{Department of Physics and Astronomy and California NanoSystems Institute, University of California, Los Angeles, Los Angeles, CA 90095, USA}

\begin{abstract}
Activating metamagnetic transitions between ordered states in van der Waals magnets and devices bring great opportunities in spintronics. We show that external pressure, which enhances the interlayer hopping without introducing chemical disorders, triggers multiple metamagnetic transitions upon cooling in the topological van der Waals magnets Mn(Bi$_{1-x}$Sb$_x$)$_4$Te$_7$, where the antiferromagnetic interlayer superexchange coupling competes with the ferromagnetic interlayer coupling mediated by the antisite Mn spins. The temperature-pressure phase diagrams reveal that while the ordering temperature from the paramagnetic to ordered states is almost pressure-independent, the metamagnetic transitions show non-trivial pressure and temperature dependence, even re-entrance. For these highly anisotropic magnets, we attribute the former to the ordering temperature being only weakly dependent on the intralayer parameters, the latter to the parametrically different pressure and temperature dependence of the two interlayer couplings. Our independent probing of these disparate magnetic interactions paves an avenue for efficient magnetic manipulations in van der Waals magnets.

\end{abstract}
\pacs{}
\date{\today}
\maketitle
 
\section{Introduction}

    Van der Waals (vdW) magnets have laid the material foundation for engineering two-dimensional (2D) thin-film devices and heterostructures with intrinsic magnetism. Triggering metamagnetic transitions between ordered states and understanding how such manipulations are driven open up unprecedented opportunities in magneto-electronics, spintronics and topotronics\cite{novoselov20162d,gong2017discovery,deng2018gate,wang2018electric,kong2019vi3,sun2020room,may2019ferromagnetism,gati2019multiple,li2021van,huang2017layer,huang2018electrical,jiang2018controlling,jiang2018electric,webster2018strain,wu2019strain,song2019switching}. Versatile means, including layer-thickness engineering, electro-gating, chemical doping, strain, pressure, etc. have been actively explored to modify the three major interactions including magnetic anisotropy, interlayer and intralayer magnetic couplings, with the aim of tuning the competing magnetic states. 
    However, due to the lack of vdW magnets with comparable ferromagnetic (FM) and antiferromagnetic (AFM) energies, despite extensive efforts, the activation of the metamagnetic transitions between these two states has only been unambiguously experimentally realized in CrI$_3$ and CrSBr insulators \cite{huang2017layer,huang2018electrical,jiang2018controlling,jiang2018electric,wu2019strain,song2019switching,webster2018strain,cenker2022reversible,wilson2021interlayer}, and there they are likely triggered by
    the structural changes. This has hindered progress in understanding the
roles that these disparate magnetic interactions play in driving such transitions.

 \begin{figure*}
 \centering
  \includegraphics[width=5in]{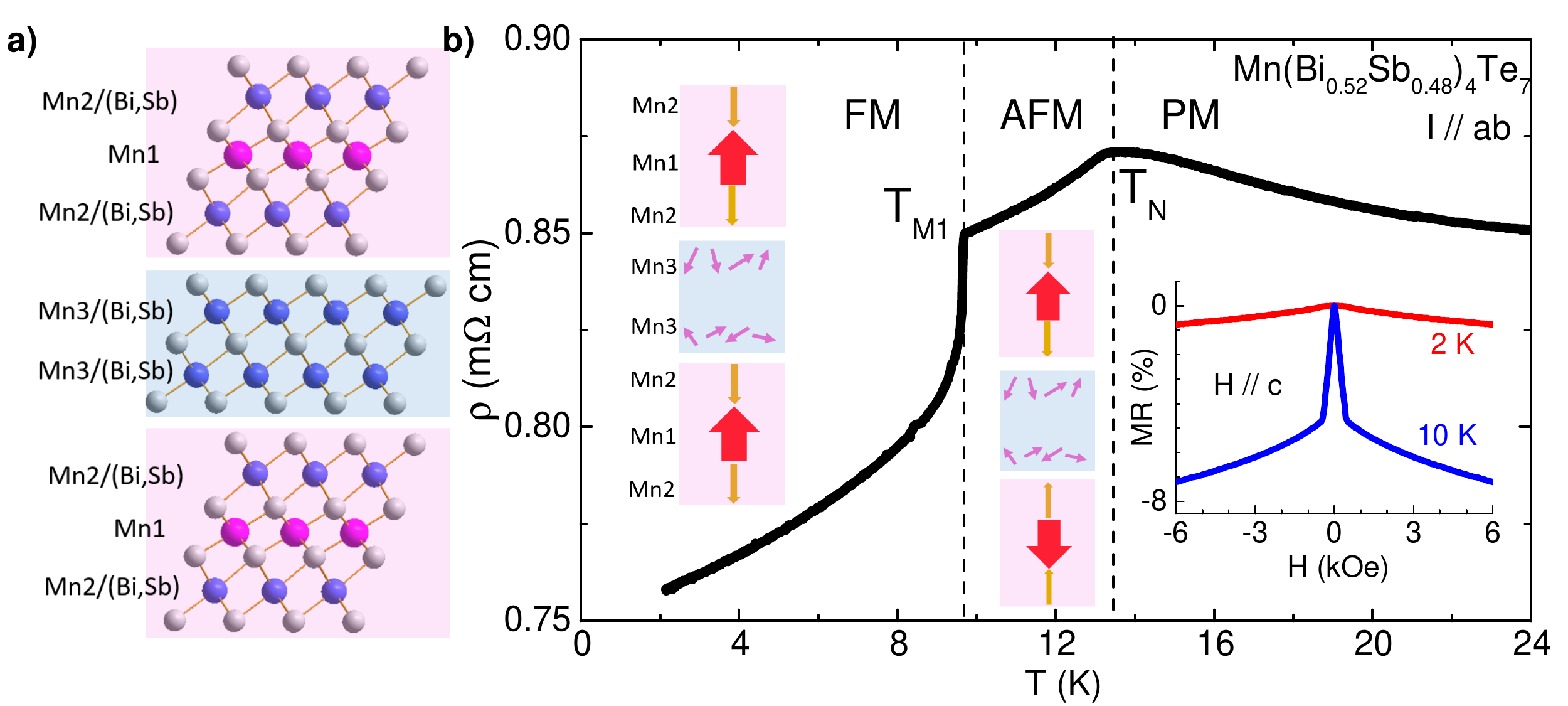}
  \caption{(a): The crystal structure of Mn(Bi$_{1-x}$Sb$_x$)$_4$Te$_7$. Sb doping can introduce Mn$_{\rm{(Bi, Sb)}}$ antisites. Mn1 represents the Mn atoms on the Mn site; Mn2 labels the Mn atoms on the (Bi, Sb) site in the SLs; Mn3 denotes the Mn atoms on the (Bi, Sb) site in the QLs. (b) The temperature-dependent resistivity, $\rho(T)$ of $x=0.48$ sample under ambient pressure, with schematics of the magnetic structures \cite{hu2021tuning}. Upon cooling, Mn1 sublattice undergoes PM $\rightarrow$ AFM $\rightarrow$ FM transitions. In these ordered states, Mn1 and Mn2 sublattices are always AFM to each other along the $c$ axis while Mn3 spins are paramagnetic. Right inset: its magnetoresistance MR$(H)$ in the FM state (2 K) and AFM state (10 K).}
  \label{RT}
\end{figure*}

%Comparing to  
 Recently, the MnBi$_{2n}$Te$_{3n+1}$ (MBT) family has been discovered to be intrinsic vdW magnets with non-trivial band topology \cite{lee2013crystal,otrokov2019prediction,zhang2019topological,li2019intrinsic,aliev2019novel,otrokov2019unique,147,wu2019natural,1813,klimovskikh2020tunable,ding2020crystal,shi2019magnetic,chen2019topological,lee2019spin,tian2019magnetic,gordon2019strongly}. They are composed of alternating $(n-1)$ [Bi$_2$Te$_3$] quintuple layers (QLs) and one [MnBi$_2$Te$_4$] septuple layer (SL). In the 2D limit of MnBi$_2$Te$_4$, due to the interplay of magnetism and band topology, emergent phenomena including the quantized Hall conductance, Chern insulator state and large layer Hall effect have been observed \cite{deng2020quantum,liu2020robust, ge2020high,gao2021layer}. Besides their fascinating non-trivial band topology, this is a family with great structural and magnetic tunability. With increasing $n$, MnBi$_2$Te$_4$, MnBi$_4$Te$_7$ and MnBi$_6$Te$_{10}$ become A-type antiferromagnets, while MnBi$_8$Te$_{13}$ becomes FM. Particularly, the as-grown MnBi$_4$Te$_7$ and MnBi$_6$Te$_{10}$ may become FM under certain growth conditions \cite{wu2019natural,yan2021delicate}, indicating close proximity of FM and AFM energy scales in this family. Chemical doping has been used to tune the magnetism in MBT \cite{chen2019intrinsic,yan2019evolution,hu2021tuning,liu2021site,xie2021charge}. The effect can be best seen in Mn(Bi$_{1-x}$Sb$_{x}$)$_4$Te$_7$, where a doping-dependent metamagnetic transition between the FM and AFM states is observed upon cooling \cite{hu2021tuning}. However, the vacancies and antisite disorder introduced by doping are uncontrollable, making the delineation of the effects of the three major magnetic interactions challenging. 
 On the other hand, %although the external 
external pressure serves as a semi-in-situ tuning knob and compresses the unit cell without changing chemical disorder. This tuning knob becomes even more advantageous for the vdW materials since it compresses more along the out-of-plane direction than in the plane \cite{zhang2021pressure, chen2019suppression,pei2020pressure,pei2022pressure}. This may lead to a much stronger pressure effect on the interlayer coupling than the intralayer parameters. However, it has been seldom applied to the MBT family where the understanding of these magnetic interactions still remain elusive \cite{chen2019suppression,pei2022pressure,shao2021pressure,pei2020pressure}. 
 Here, using electrical transport and magnetometry measurements, we show the extremely sensitive pressure-driven activation and manipulation of the metamagnetic transitions in Sb-doped MnBi$_4$Te$_7$, which are non-trivial and even re-entrant. By systematic investigations of the temperature-pressure ($T-P$) phase diagrams, we further demonstrate that our experiment provides a rare platform to distinguish and understand the effects of magnetic anisotropy, interlayer and intralayer couplings on vdW magnetism.

\section{Results}
In Mn(Bi$_{1-x}$Sb$_{x}$)$_4$Te$_7$, Sb atoms not only replace Bi atoms but also promote site mixing between Sb and Mn. The presence of Mn$_{\rm{(Bi, Sb)}}$ antisites leads to Mn1, Mn2 and Mn3 sublattices, as depicted in Fig. 1 (a). As a consequence of the competition between the ``standard" Mn1-Mn1 AFM interlayer superexchange interaction assisted by hopping to the Bi/Sb/Te atoms and the Mn3-mediated Mn1-Mn1 FM interlayer exchange interaction, this series has comparable FM and AFM energies \cite{hu2021tuning}. The $x=0$ member is an A-type antiferromagnet with the spins ferromagnetically aligned in the $ab$ plane but antiferromagnetically aligned along the $c$ axis. For $x < 0.57$, the dominant Mn1 sublattice first enters the A-type AFM state and then undergoes an AFM to low-temperature FM metamagnetic transition. For $x\ge 0.57$, only one transition from PM to FM state exists. 

In MBT, magnetism strongly couples with the charge carriers. As an example, Fig. \ref{RT} (b) presents the temperature-dependent resistivity, $\rho(T)$, magnetoresistance (MR) and the cartoon magnetic structures of Mn(Bi$_{0.52}$Sb$_{0.48}$)$_4$Te$_7$ \cite{hu2021tuning}. Note that since the Mn2 spins are always antiferromagnetically coupled with the Mn1 sublattice, we will, for simplicity, focus on the Mn1 sublattice. Upon cooling, $\rho(T)$ decreases with two slope changes. One is at $T_N$ = 13.3 K, associated with the PM to AFM transition of the Mn1 lattice.  The other is a sharp resistivity drop at $T_{M1}$ = 9.7 K, arising from the AFM to FM metamagnetic transition of the Mn1 spins. MR$(H)$ with $H // c$ and $I //ab$ shows distinct behaviors when the Mn1 spins are in the FM and AFM states, as seen in the inset of Fig. \ref{RT} (b). At 10 K, a sharp decrease of MR appears at around 0.4 kOe by $\sim 5\%$. This drop comes from the decrease of the spin disorder scattering when the system goes from the AFM to forced FM state at 0.4 kOe, the spin-flip field of Mn1 \cite{hu2021tuning}. On the other hand, at 2 K, where the Mn1 spins are in the FM state, a weak monotonic decrease in MR occurs across the coercive field. In this paper, together with the magnetic data, the distinct MR behaviors discussed above will be used to differentiate if the Mn1 spins are AFM or FM.

 \begin{figure*}
 \centering
\includegraphics[width=\textwidth]{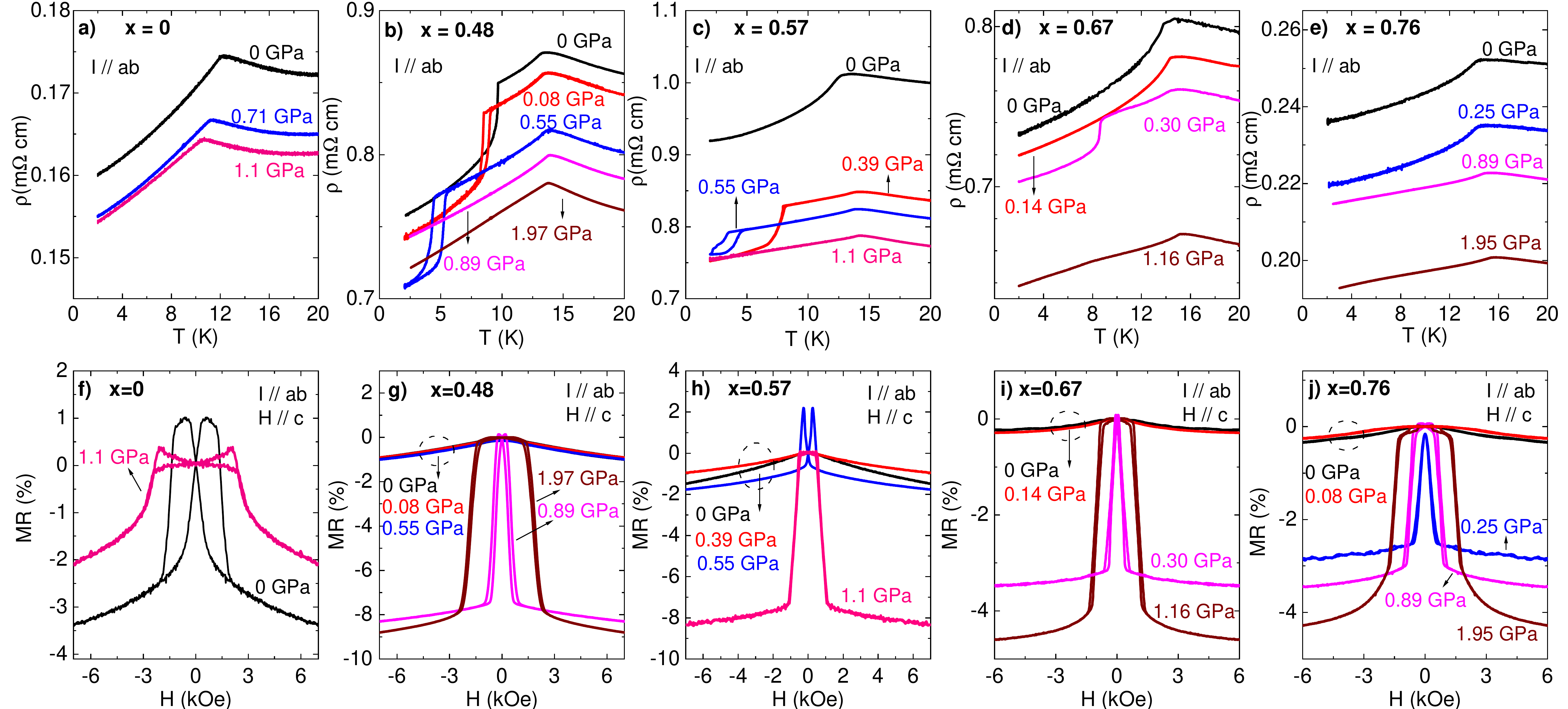}
  \caption{The effect of external pressures on the electrical properties of Mn(Bi$_{1-x}$Sb$_x$)$_4$Te$_7$ ($x=0$, 0.48, 0.57, 0.67 and 0.76). (a)-(e) $\rho(T)$ at different pressures with the current $I//ab$ plane. (f)-(j): MR$(H)$ at 2 K under pressures with $I//ab$ and $H //c$. The sharp drop in MR indicates the Mn1 spins are at the AFM state while the MR showing weak field dependence suggests FM state of the Mn1 spins. }
  \label{RT-MR}
\end{figure*}

\subsection{The pressure-switching of magnetic states }

Five different doping levels ($x=0, 0.48, 0.57, 0.67$ and $0.76$) were selected for the pressure study, as shown in Fig. \ref{RT-MR}. For $x=0$, at 0 GPa, Mn(Bi$_{1-x}$Sb$_x$)$_4$Te$_7$ is AFM below $T_N=12.2$ K, as indicated by the anomaly in $\rho(T)$ (Fig. \ref{RT-MR} (a)) and the sharp decrease of MR at the spin-flip field of $\sim$ 2 kOe at 2 K (Fig. \ref{RT-MR} (f)). $T_N$ decreases slightly with pressure, similar to a previous study \cite{shao2021pressure}. All MR curves at $P\leq1.1$ GPa show a clear sudden drop, indicating that the system remains in the AFM state under pressure.

For $x=0.48$ (Fig. \ref{RT-MR} (b) and (g)), the ordering temperature $T_N$ is essentially unaffected by pressure up to 1.97 GPa, the highest pressure we applied. In sharp contrast, the FM to AFM metamagnetic transition at $T_{M1}$ is extremely sensitive to pressure and becomes first-order like under pressure. At 0.89 GPa, $T_{M1}$ is completely suppressed, leaving the ground state as AFM. Indeed, our MR data at 2 K (Fig. \ref{RT-MR}(g)) shows a weak monotonic decrease for $P \leq 0.55$ GPa that is consistent with a FM ground state, while for $P\geq 0.89$ GPa, the sharp drop in MR reveals the AFM ground state. 

For $x=0.57$, at 0 GPa, the PM to FM transition is revealed by the single resistivity anomaly at $T_C$ shown in Fig. \ref{RT-MR}(c) and the weak monotonic MR decrease at 2 K (Fig. \ref{RT-MR}(h)). The application of hydrostatic pressure has a dramatic effect on this material. At 0.39 GPa, $\rho(T)$ shows two slope changes, similar to what we observed for $x=0.48$ below 0.55 GPa, suggesting the emergence of a pressure-induced AFM phase between $T_N=14.4$ K and $T_{M1}=8.0$ K. $T_N$ slightly increases with pressure while $T_{M1}$ is completely suppressed above 0.55 GPa. MR measurements confirm that the ground state is AFM for $P \leq 0.55$ GPa and FM for $P\geq1.1$ GPa.

For $x=0.67$, although the pressure effect seems similar to that of %the
$x=0.57$, %one
 the ground state at 0.30 GPa is a puzzle. The $\rho(T)$ data implies PM $\rightarrow$ AFM $\rightarrow$ FM transitions and thus a FM ground state (Fig. \ref{RT-MR}(d)); however, the MR at 2 K shows a sharp drop by 3\%, indicating an AFM ground state (Fig. \ref{RT-MR}(i)). These contradicting observations may suggest the re-entrance of AFM state, which will be discussed later.

The $x=0.76$ data are even more intriguing. Despite the envelope of $\rho(T)$ barely changing under pressure (Fig. \ref{RT-MR}(e)), remarkably, the MR data shown in Fig. \ref{RT-MR}(j) suggests that this compound is the most sensitive %one 
to %the 
pressure among all, with the ground state being FM below 0.08 GPa and AFM above 0.25 GPa. 

\begin{figure*}
 \centering
  \includegraphics[width=7.1
 in]{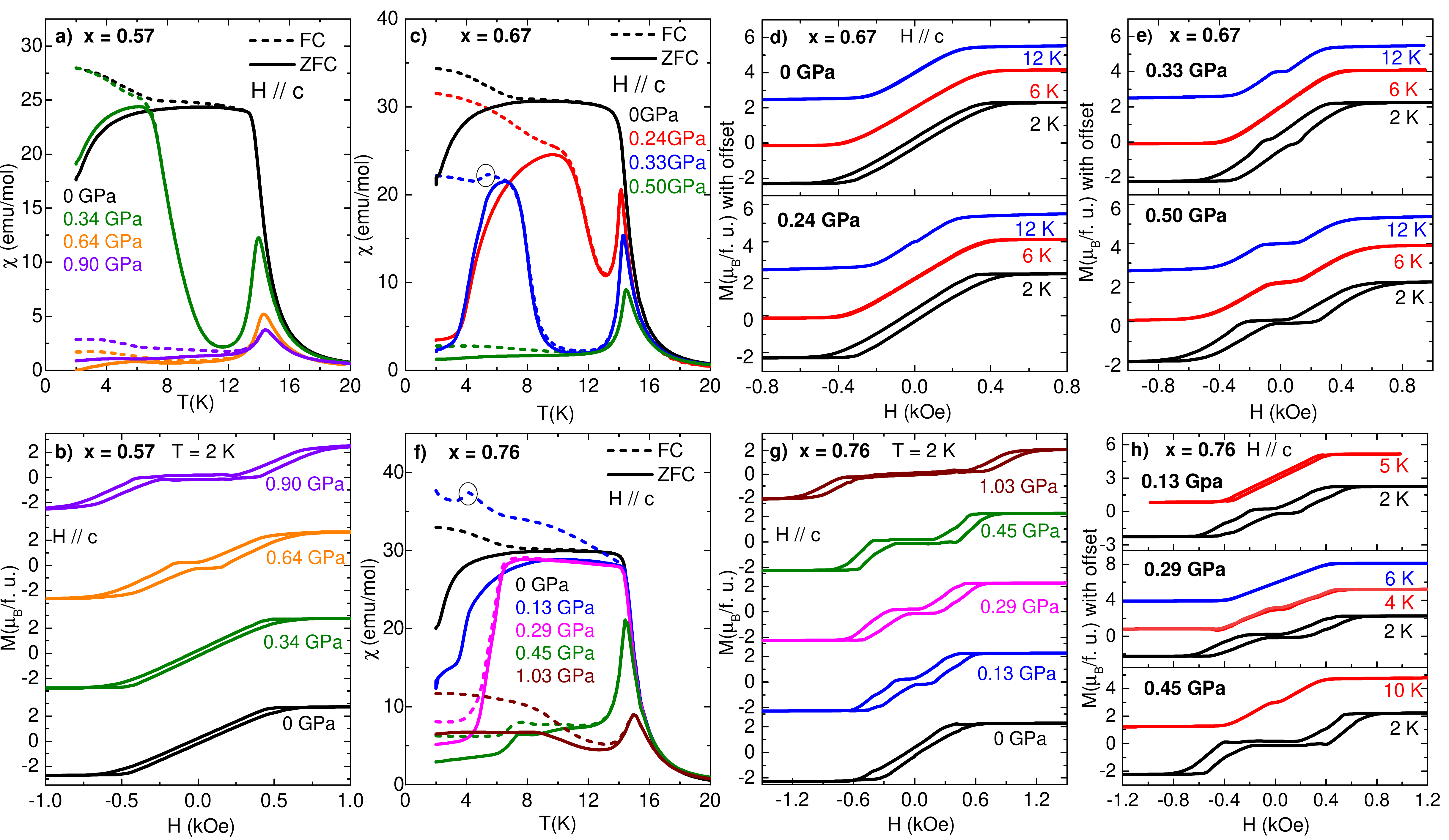}
  \caption{The effect of external pressures on the magnetic properties of Mn(Bi$_{1-x}$Sb$_x$)$_4$Te$_7$ ($x=0.57$, 0.67 and 0.76): (a), (c), (f) The temperature dependant ZFC and FC magnetic susceptibility $\chi(T)$ with $H//c$. (b), (g): The isothermal magnetization $M(H)$ at 2 K at different pressures for $x=0.57$ (b), and 0.76 (g). (d), (e) and (h): at fixed pressures, the $M(H)$ curves at different temperatures for $x = 0.67$ ((d) and (e)), and $x=0.76$ (h). $M(H)$ data are analyzed to remove the lead signal, so slight discontinuity in data are induced.}
  \label{MTMH}
\end{figure*}

To further investigate the aforementioned pressure-induced metamagnetic transitions and the puzzling ground states, magnetic susceptibility $\chi(T)$ and isothermal magnetization $M(H)$ were measured for the $x=0.57$, 0.67 and 0.76 compounds with $H ||c$, as shown in Fig. \ref{MTMH}. For $x = 0.57$, at ambient pressure, $\chi(T)$ shows a steep upturn plateau at $T_C$ of 13.4 K and a large bifurcation between the field-cooled (FC) and zero-field-cooled (ZFC) data (Fig. \ref{MTMH}(a)); together with the typical hysteresis loop in $M(H)$ at 2 K (Fig. \ref{MTMH}(b)), this indicates a FM ground state. At 0.34 GPa, the envelope of $\chi(T)$ evolves to show two anomalies. A cusp (PM to AFM) at $T_N = 14.0$ K and a steep upturn plateau (AFM to FM) at $T_{M1} = 6.7$ K. Under higher pressures, $T_N$ slightly moves to higher temperature, while $T_{M1}$ is completely suppressed above 0.64 GPa, where the $M(H)$ at 2 K shows the spin-flip feature, consistent with an AFM ground state. 

For $x=0.67$, $\chi(T)$ resembles that of $x=0.57$. Under pressure, a cusp feature, absent at ambient pressure, sets in around 14 K, showing weak pressure dependence; meanwhile, the steep upturn feature with a broad maximum is suppressed by pressure. However, closer examination of the 0.33 GPa data reveals that upon cooling, following the broad maximum, an additional ``kink" feature highlighted by the circle in Fig. \ref{MTMH}(c) emerges in the FC $\chi(T)$, suggesting three sequential magnetic transitions. Indeed, the $M(H)$ data shown in Figs. \ref{MTMH}(d-e) indicate complex phase transitions. At 0 GPa, the sample remains FM below $T_C$; at 0.24 GPa, it undergoes a PM $\rightarrow$ AFM $\rightarrow$ FM transition; at 0.33 GPa, it goes from PM to AFM at $T_N$, then the first metamagnetic transition from the AFM state to the FM state at $T_{M1}$, and then the second metamagnetic transition from the FM state to the AFM state at $T_{M2}$. Above 0.50 GPa, it remains AFM below $T_N$. 

The $x=0.76$ data show some similarities with those for $x=0.67$. At 0.13 GPa, following the broad maximum (Fig. \ref{MTMH}(f)), a kink feature in FC $\chi(T)$ sets in at $T_{M2} = 4.2$ K which, together with Fig. \ref{MTMH}(g), suggests a metamagnetic transition from a FM state to an AFM state at $T_{M2}$. $T_{M2}$ increases under pressure, which manifests as a sharp drop in $\chi(T)$ at 0.29 GPa and then becomes a cusp feature at 0.45 GPa. The sequence of these phase transitions can also be inferred from the $M(H)$ data (Fig. \ref{MTMH}(h)). Upon cooling, at 0 GPa, it stays FM; at 0.13 and 0.29 GPa, it is FM $\rightarrow$ AFM; above 0.45 GPa, it remains AFM. 

 \begin{figure*}
 \centering
  \includegraphics[width=7in]{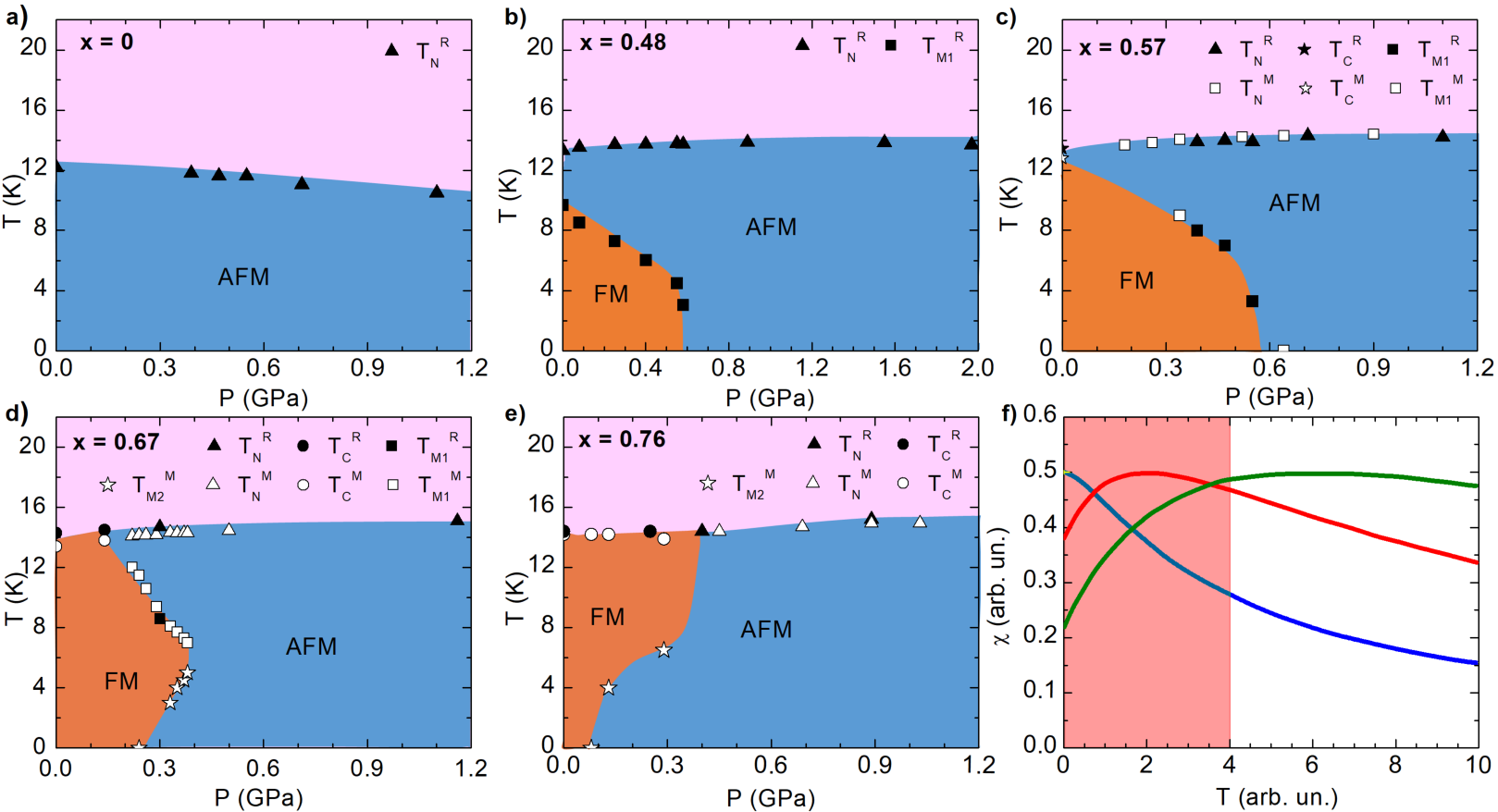}
  \caption{(a)-(e) Temperature-pressure ($T-P$) phase diagrams for Mn(Bi$_{1-x}$Sb$_x$)$_4$Te$_7$ ($x=0$, 0.48, 0.57, 0.67 and 0.76) under pressures. $T_C$ and $T_N$ are the magnetic ordering temperatures of a PM to FM or AFM transition, respectively. $T_{M1}$ and $T_{M2}$ are the metamagnetic temperatures where an AFM $\rightarrow$ FM transition or a FM $\rightarrow$ AFM transition appears upon cooling, respectively. $T_N^R$, $T_{C}^R$ and $T_{M1}^R$ were extracted by taking the first derivative of resistivity data. $T_N^M$, $T_C^M$, $T_{M1}^M$ and $T_{M2}^M$ are determined by the first derivative of magnetic susceptibility data and then confirmed with isothermal magnetization measurements across critical temperatures. Note: the 
  $\square/\blacksquare  $
  phase line representing the AFM to FM metamagnetic transition upon cooling is a first-order phase line while the others are all second-order phase lines. (f) Typical behavior of susceptibility in a system with antiferromagnetic or spin-glass type correlations. The blue/red/green lines correspond to $T_{CW}\approx 3, 7$ and 15. note that in a limited temperature range shown by pink shading, 
  $\chi$ can either grow, decay, or show non-monotonic behavior.  }
  \label{3Dpd}
\end{figure*}

\subsection{Doping-dependent $T-P$ phase diagrams}
Based on our data in Fig. 2 and 3, Figs. \ref{3Dpd} (a)-(e) summarize the temperature-pressure ($T-P$) phase diagrams. The upper phase line represents the ordering transition from the PM state to the ordered state while the lower phase line marks the metamagnetic transitions between ordered states. Results from resistivity and magnetic data are in excellent agreement with each other. 

We found that the upper phase line is weakly pressure dependent. In stark contrast, the lower phase line is unusually sensitive to pressure. Overall, the external pressure squeezes out FM and stabilizes AFM, leading to the unconventional lower phase line, which manifests in three remarkable manners. For the lower doping levels ($x=0.48$ and 0.57), $T_{M1}$, which marks the first-order AFM to FM metamagnetic transition, monotonically decreases with external pressure. For the intermediate doping level of $x=0.67$, the re-entrant AFM state appears at $T_{M2}$ from $\sim 0.2$ GPa to 0.4 GPa, leading to a non-monotonic phase line with the first-order phase line of $T_{M1}$ decreasing with pressure and the second-order phase line of $T_{M2}$ increasing with pressure. For the high doping level $x=0.76$, $T_{M1}$ disappears, but $T_{M2}$ monotonically increases with pressure and finally intersects with the upper phase line.

We also note that the pressure required to switch the ground state from FM to AFM decreases with $x$, ranging from $\sim$ 0.6 GPa at $x=0.48$ to $\sim$ 0.1 GPa at $x=0.76$.

\subsection{Microscopic understanding of the $T-P$ phase diagrams}

The fact that these phase diagrams are strongly doping-dependent suggests that a model that considers the chemical complexity in this series has to be developed for quantitative understanding. 

Several aspects are particularly unexpected, if not counterintuitive. First, as represented by the upper phase line, at sufficiently high doping levels (Fig. \ref{3Dpd}d,e), the magnetic order switches suddenly from FM to AFM with pressure, yet the transition temperature
($T_C$ or $T_N$, respectively)  is basically unchanged (a very tiny notch is 
barely discernible at the triple point at $x=0.76$). Na\"ively speaking, one would expect that at the triple point where the magnetic order is fully frustrated, the transition temperature, if at all existed, should be much lower.

Second, pressure induces ferromagnetism at 
low temperature, but the order switches back to AFM upon cooling. Such FM to AFM metamagnetic transitions are rare (the best known example is FeRh) \cite{vinokurova1976pressure, stern2017giant}, and usually driven by the large volume effect at the metamagnetic transition (Clausius-Clapeyron theorem). This does 
not seem to be the case in our material, 
especially in view of the fact that the 
metamagnetic transitions marked with stars
in Fig. \ref{3Dpd}d,e are second-order and are unlikely to be due to the volume effect.

Finally, the {\it sign} of the pressure coefficient of the metamagnetic transitions, $dT_M/dP$, varies with
doping: $dT_M/dP<0$ for $x\alt 0.6$, $dT_M/dP>0$ for $x\agt 0.7$, and for 
$x=0.67$ it is positive at low,  and negative at higher temperatures.

These seemingly perplexing observations are all rooted in the unique 
separation of magnetic interaction in Mn(Bi$_{1-x}$Sb$_x$)$_4$Te$_7$. Indeed, 
it can be viewed, in a first approximation, as two overlapping magnetic subsystems, shaded in Fig. \ref{RT}a as pink and blue. In the discussion below, for simplicity,
the former shall be referred to %shall refer to the former
as Mn1 and the latter as Mn3.

Let us first consider Mn1. A single Mn1 layer forms a 2D magnetic system with strong FM intraplanar coupling $J>0$ and weak interplanar coupling of varying 
sign $|J_\perp|\ll J$. In addition, it may have intraplanar magnetic
anisotropy, which can, without loss of generality, be absorbed into a single-site term, so that the total magnetic Hamiltonian looks like
\begin{equation}
    H_{11}=\sum_{ii'}J{\bf{S_i}\cdot{S_{i'}}}+\sum_{ij}J_\perp{\bf{S_i}\cdot{S_j}}+\sum_{i}K{S_{iz}}^2.
\label{a}
\end{equation}
 Here $i,i'$ denote sites in the same layer, and $i,j$ in the neighboring layers.
 
 \subsubsection{Weakly pressure-dependent ordering temperatures}

Per Mermin-Wagner theorem, in the 2D limit, a system does not order at any finite
$T$ if the exchange coupling is isotropic or with an easy-plane anisotropy ($K>0$). On the other hand, a 2D easy-axis system with $K<0$ and/or nonzero $J_\perp$ orders at a
temperature generally determined as
\begin{equation}
T_{c}=\frac{a|J|}{b+\log(J/J_{\rm {eff}})}%
\label{1}
\end{equation}
where $a$, $b$ and $c$ are constants
of the order of 1, and ${J_{\rm {eff}}}$ is a combination of $J_{\perp}$ and $K$ that reduces to K in the $J_{\perp}\rightarrow 0$ limit. For instance, in Ref.\cite{andrei2007magnetic} an expression for ${J_{\rm {eff}}}$
was derived as $J_{\rm{eff}}=K+J_\perp+\sqrt{K^2+2KJ_{\perp}}$.

According to the Stoner-Wohlfarth model\cite{stoner1948mechanism}, we can estimate $|K|$ and $|J_{\perp}|$. For the $x=0$ sample with the spin-flip transition, $|SJ_{\perp}|=g\mu_BH_c/z$ and $|SK|=g\mu_B(H_{ab}-2H_c)/2$, where $H_{ab}=1.2$ T is the field for the Mn1 spins to saturate along the $ab$ plane and $H_c=0.14$ T is the spin-flip field along the $c$ axis, $g=2$, $S=5/2$ and $z=2$ is the number of nearest Mn interlayer neighbors. The calculated $|SK|$ = 0.053 meV and $|SJ_{\perp}|=0.008$ meV. So, we conclude that our materials are in the regime where $K^2\gg J_{\perp}^2.$ In this regime, the ordering 
temperature ($T_N$ or $T_C$) can be approximated as
\begin{equation}
T_{c}\approx \frac{a|J|}{b+\log(J/K)}
\label{1a}
\end{equation}
As we can see, $T_c$ is only controlled by $J$ and $K$, which characterize the intralayer magnetic dynamics. Specifically, $T_c$ depends logarithmically weakly on the ratio of $J/K$. So, the ordering temperature is expected to be weakly pressure dependent, with maybe a tiny notch
right at the triple point where $J_{\perp}$ is fully compensated. %Just to give an idea of the scale, if $|J|=100$ and $K=1$, $T_{c}$ changes by around 0.4\% between $J_{\perp}=0$ and $J_{\perp}=0.2K.$ 
Indeed, in our experiment, $T_N$ or $T_C$ varies little in the entire set of experiments, between $\sim 11$ K ($x=0$, $P=1.1$ GPa) and $\sim 15$ K ($x=0.76$, $P=0.9$ GPa). 

Therefore, as a material system in the $K^2\gg J_{\perp}^2$ regime, although the sign of $J_{\perp}$ defines the long-range order in the $c$ direction, the ordering temperature depends logarithmically weakly on the ratio of $J/K$. Consequently, neither $T_N$ nor $T_C$ are sensitive to the external pressures we applied (only up to 2 GPa) and are oblivious to the metamagnetic transitions under pressure.

 \subsubsection{Pressure-activation of metamagnetic transitions}

 Let us now turn to the pressure dependence of the metamagnetic transition. To understand it, we observe that a FM ordered Mn1 plane induces an exchange 
bias field $H_{ex}$ in a magnetically disordered Mn3 layer. Assuming that the magnetic susceptibility
of the latter is $\chi_3(T)$, we can add the fourth term to the Hamiltonian in Eq. \ref{a}
, namely
\begin{equation}
H_{13}=\pm\chi_3(T)H_{ex}^2,
\label{Hex}
\end{equation} where the plus sign corresponds to
AFM stacking of the Mn1 plane, and the minus to the FM stacking. Obviously, this
Mn3-mediated interaction is always ferromagnetic, and competes with the 
standard Mn1-Mn1 superexchange \cite{hu2021tuning}.

Let us now estimate the pressure dependence of both interlayer terms using a simple Hubbard model. To this end, we consider two possible exchange paths. One
is the ``standard'' superexchange, when an electron virtually hops from the effective  Mn1 layer (which includes the entire pink region in Fig. \ref{RT}a) to anions in the effective Mn3 layer (Bi, Sb or Te), and then to the next 
Mn1 layer. We will assign to Mn one effective $d$ level, $E_d,$ and to all anions
one effective $p$ level, $E_p,$ with the charge transfer energy $\Delta E=E_d-E_p$, and a Hubbard repulsion $U$.
The second exchange path is from the effective Mn1 to individual Mn3 ions.
Importantly, $E_p,\ E_d$ and $U$ are atomic parameters and are not sensitive to pressure.
On the contrary, the hopping amplitudes, $t_{pd}$ for the former path and 
 $t_{dd}$ for the latter are very sensitive to the interlayer distance.

The first path defines the standard AFM superexchange, 
\begin{equation}
J_{\perp}^{\rm afm}\propto t_{pd}^4/\Delta E^2U. \label{afm}
\end{equation}
The second determines the exchange bias parameter in Eq. \ref{Hex},
\begin{equation}
H_{ex}\propto t_{dd}.
\end{equation}
Per Eq. \ref{Hex}, this generates \begin{equation}
J_{\perp}^{\rm fm}\propto t_{dd}^2\chi_3(T).\label{fm}
\end{equation}
Note that the same conclusion can be achieved by diagonalizing a three site 
Hubbard model with half-filling under an assumption that all sites have the same
$U$ and hopping $t_{12}=t_{13}=t_{dd}$.

The corollary is that, while both $J_{\perp}^{\rm afm}$ and $J_{\perp}^{\rm fm}$ are expected to increase with pressure, the former grows as the forth power
of the effective hopping, and the latter only as the second power. Since the metamagnetic temperature ($T_{M1}$ or $T_{M2}$) is defined by $J_\perp$ (the transition 
occurs when it is fully compensated, $J_\perp=0$), it is very sensitive to pressure because the AFM part grows much 
faster with pressure. Therefore, as expected from the model analysis above, pressure favors the AFM order.

 \subsubsection{Exotic temperature dependence
of metamagnetic transitions}

The most intriguing part is the exotic temperature dependence
of the metamagnetic transition, with at least one concentration ($x=0.67$) with the recurrent
behavior. In order to understand that, recall that the only temperature-dependent parameter in Eqs. \ref{afm} and \ref{fm} is $\chi_3(T)$.
Let us estimate its temperature dependence. For small
doping levels we can neglect the internal interactions inside the Mn3 layer, so
that $\chi_{3}(T)$ is described by the Curie law, $\chi_{3}(T)\propto1/T.$ At
higher concentrations it is reasonable to assume (and this is also supported
by neutron data) that the interactions in the Mn3 layer are random in sign and
amplitude \cite{hu2021tuning}, so on the mean field level they would freeze into a spin glass state with
the net magnetization $\left\langle M_{3}\right\rangle =0.$ While fluctuations beyond the mean field can completely destroy the spin-glass transition, 
or drive it to extremely low temperatures, the
magnetic susceptibility of such systems would behave similar to that in a usual antiferromagnet. That is to say, the susceptibility will decay
at high $T$ as $\chi_{3}(T)\propto1/(T+T_{CW})$ (we are using the convention where
$T_{CW}>0$ for the antiferromagnetic response), and has a maximum at some
temperature $T_{0}<T_{CW},$ and $T_{CW}$ is on the order of the average
interaction strength in the Mn3 plane. A typical $\chi_{3}(T)$ is shown in
Fig. \ref{3Dpd}(f), where the approximate behavior of $\chi_{3}(T)$ for $T_{CW}=3,$ $7$ and $15$ is plotted (in arbitrary units).
Note that $T_{CW}$ is growing with the concentration, first very weakly, than
rapidly. Therefore the temperature range of interest (the shaded region in 
Fig. \ref{3Dpd}) may fall either entirely in the range of $\chi_{3}(T)$
decreasing with temperature (small concentration of Mn3), or entirely in the
range of increasing $\chi_{3}(T)$ (large concentrations), or even span both
regimes (intermediate concentrations). 

Now we can complete the microscopic explanation of all nontrivial behaviors listed in the beginning of this section. For $x=0$, the Mn3 concentration is very low, so $J_{\perp}^{\rm afm} > J_{\perp}^{\rm fm}$ holds below $T_N$ for all pressures. For $x=0.48$ and 0.57, the relatively larger Mn3 concentration (still low concentration case) makes $J_{\perp}^{\rm fm}$ strong enough to compete with $J_{\perp}^{\rm afm}$. Since $\chi_3$ and thus $J_{\perp}^{\rm fm}$ increase upon cooling, below the ordering temperature, an AFM to FM transition appears at $T_{M1}$. Upon compressing, $J_{\perp}^{\rm afm}$ increases faster than $J_{\perp}^{\rm fm}$, so $T_{M1}$ decreases with pressure. 
For $x=0.67$ (intermediate concentration case), at low pressures, $J_{\perp}^{\rm fm} > J_{\perp}^{\rm afm}$ holds for $T < T_C$. But above a threshold pressure, $J_{\perp}^{\rm afm}$ is favored over $J_{\perp}^{\rm fm}$ at $T_N$. Then upon cooling, $\chi_3$ and $J_{\perp}^{\rm fm}$ first rise and then decrease, so an AFM to FM transition appears at $T_{M1}$ and then a FM to AFM transition shows up at $T_{M2}$. Finally, with increasing pressure, the faster growing $J_{\perp}^{\rm afm}$ leads to decreasing $T_{M1}$ and increasing $T_{M2}$. For $x=0.76$, $\chi_3$ decreases upon cooling (large concentration case), as well as $J_{\perp}^{\rm fm}$. Therefore the FM to AFM transition occurs at $T_{M2} < T_C$. With increasing pressure, $J_{\perp}^{\rm afm}$ grows faster, thus $T_{M2}$ increases with pressure.

\section{Discussion}
In the topological vdW magnets Sb-doped MnBi$_4$Te$_7$, the Mn spins have the stacking of -(Mn1+Mn2)-Mn3-(Mn1+Mn2)-. This arrangement leads to two types of interlayer exchange pathways, the antiferromagnetic superexchange interaction through anions and the ferromagnetic exchange interaction mediated by Mn3, which have made this material system superior for tuning. Through the application of external pressure, we observe rare pressure-activated metamagnetic transitions with non-trivial pressure- and temperature- dependence and even re-entrance. Particularly, for the $x=0.67$ sample with a PM to FM transition at ambient pressure, it undergoes sequential transitions from PM $\rightarrow$ AFM $\rightarrow$ FM $\rightarrow$ AFM at 0.33 GPa. These unconventional behaviors are shown to be due to the competition between the aforementioned two interlayer interactions since the former is the forth power of the effective interlayer hopping with weak temperature-dependence and the latter is the second power,
and has strong temperature dependence. Furthermore, in stark contrast, the pressure effect on the ordering temperature from the PM to ordered state, whether $T_N$ or $T_C$, is weak. We argue that this is because the ordering temperature depends logarithmically weakly on the ratio of the intralayer coupling and magnetic anisotropy for vdW magnets with strong magnetic anisotropy, the case here.

Our results have several implications. First, by pressurizing Sb-doped MnBi$_4$Te$_7$, we demonstrate the independent probing of magnetic anisotropy, intralayer and interlayer magnetic couplings in vdW magnets. We elucidate their distinct pressure dependence and reveal their individual effects on the ordering temperatures and the triggering of the metamagnetic transitions under pressure and temperature. This is crucial for the applications of vdW magnets and devices in magneto-electronics, spintronics and topotronics. Third, the argument that Mn3-mediated interlayer exchange interactions must be FM in the Mn-Mn3-Mn stacking, points out a material design strategy to engineer FM couplings in vdW and heterostructural magnets. Lastly, as vdW magnets with non-trivial band topology, these pressure-driven metamagnetic transitions, will surely cause pressure-driven topological phase transitions in Sb-doped MnBi$_4$Te$_7$, leading to new magnetic topological phases, which will be interesting for future investigations.

\section{Materials and Methods}

\subsection{Single crystal growth and characterization}
Single crystals were grown using the flux method \cite{hu2021tuning}. Powder X-ray diffraction was performed via a PANalytical Empyrean diffractometer (Cu K$\alpha$ radiation) to determine the phase purity. The Sb doping levels of the Mn(Bi$_{1-x}$Sb$_x$)$_4$Te$_7$ crystals refer to the actual concentrations, determined by wavelength dispersive spectroscopy.

\subsection{Magnetic and electrical property measurements}
Magnetic susceptibility and isothermal magnetization measurements under pressure were performed in a Quantum Design (QD) Magnetic Property Measurement System (MPMS3) with $H //c$. Electrical transport measurements were performed in a QD DynaCool Physical Property Measurement System (PPMS). Resistivity and magnetoresistance measurements were performed using the four-probe method, with $I//ab$ and $H//c$. To eliminate unwanted contributions from mixed transport channels of the magnetotransport data, data were collected while sweeping the magnetic field from -9 T to 9 T. The data were then symmetrized to obtain $\rho_{xx}(H)$ using $\rho_{xx}(H)=(\rho_{xx}(H)+\rho_{xx}(-H))/2$. The MR was defined as $(\rho_{xx}(H) - \rho_{xx}(0))/\rho_{xx}(0)$.

\subsection{Measurements under pressures}
For the transport properties under pressure, a C\&T Factory commercial piston pressure cell compatible with a QD PPMS was used; for the magnetic properties under pressure, a HMD pressure cell compatible with a QD MPMS3 was applied. Daphne Oil 7373 \cite{yokogawa2007solidification} was used as the hydrostatic pressure medium. A Pb piece was used as a manometer by tracking the pressure dependence of its superconducting transition, which is described by $dT_c/dP = - 0.361(5)$ K/GPa \cite{clark1978pressure}. The magnetic signal from Pb was subtracted from the total magnetic signal to obtain the magnetic data of the samples.

\subsection{Acknowledgements}

Work at UCLA was supported by the U.S. Department of Energy (DOE), Office of Science, Office of Basic Energy Sciences under Award Number DE-SC0021117. I. M. acknowledges support from DOE under the grant DE-SC0021089. C. H. thanks the support by the Julian Schwinger Fellowship at UCLA.

\medskip

%\bibliographystyle{apsrev4-1}
%\bibliography{pressure}

\begin{thebibliography}{99}


\bibitem{novoselov20162d} K. S. Novoselov, A. Mishchenko, A. Carvalho, A. H. C. Neto, 2D materials and van der Waals heterostructures. \emph{Science} \textbf{353}, aac9439 (2016).
\bibitem{gong2017discovery} C. Gong, L. Li, Z. Li, H. Ji, A. Stern, Y. Xia, T. Cao, W. Bao, C. Wang, Y. Wang, Z. Q. Qiu, R. J. Cava, S. G. Louie, J. Xia, X. Zhang, Discovery of intrinsic ferromagnetism in two-dimensional van der Waals crystals. \emph{Nature} \textbf{546}, 265-269 (2017).
\bibitem{deng2018gate} Y. Deng, Y. Yu, Y. Song, J. Zhang, N. Z. Wang, Z. Sun, Y. Yi, Y. Z. Wu, S. Wu, J. Zhu, others, Gate-tunable room-temperature ferromagnetism in two-dimensional Fe$_3$GeTe$_2$. \emph{Nature} \textbf{563}, 94-99 (2018).
\bibitem{wang2018electric} Z. Wang, T. Zhang, M. Ding, B. Dong, Y. Li, M. Chen, X. Li, J. Huang, H. Wang, X. Zhao, Y. Li, D. Li, C. Jia, L. Sun, H. Guo, Y. Ye, D. Sun, Y. Chen, T. Yang, J. Zhang, S. Ono, Z. Han, Z. Zhang, Electric-field control of magnetism in a few-layered van der Waals ferromagnetic semiconductor. \emph{Nat. Nanotechnol.} \textbf{13}, 554-559 (2018).
\bibitem{kong2019vi3} T. Kong, K. Stolze, E. I. Timmons, J. Tao, D. Ni, S. Guo, Z. Yang, R. Prozorov, R. J. Cava, VI$_{3}$-a new layered ferromagnetic semiconductor. \emph{Adv. Mater.} \textbf{31}, 1808074 (2019).
\bibitem{sun2020room} X. Sun, W. Li, X. Wang, Q. Sui, T. Zhang, Z. Wang, L. Liu, D. Li, S. Feng, S. Zhong, H. Wang, V. Bouchiat, M. Nunez Regueiro, N. Rougemaille, J. Coraux, A. Purbawati, A. Hadj-Azzem, Z. Wang, B. Dong, X. Wu, T. Yang, G. Yu, B. Wang, Z. Han, X. Han, Z. Zhang, Room temperature ferromagnetism in ultra-thin van der Waals crystals of 1T-CrTe$_2$. \emph{Nano Res.} \textbf{13}, 3358-3363 (2020).
\bibitem{may2019ferromagnetism} A. F. May, D. Ovchinnikov, Q. Zheng, R. Hermann, S. Calder, B. Huang, Z. Fei, Y. Liu, X. Xu, M. A. McGuire, Ferromagnetism near room temperature in the cleavable van der Waals crystal Fe$_{5}$GeTe$_{2}$. \emph{ACS nano} \textbf{13}, 4436-4442 (2019).
\bibitem{gati2019multiple} E. Gati, Y. Inagaki, T. Kong, R. J. Cava, Y. Furukawa, P. C. Canfield, S. L. Bud'Ko, Multiple ferromagnetic transitions and structural distortion in the van der Waals ferromagnet VI$_{3}$ at ambient and finite pressures. \emph{Phys. Rev. B} \textbf{100}, 094408 (2019).
\bibitem{li2021van} B. Li, Z. Wan, C. Wang, P. Chen, B. Huang, X. Cheng, Q. Qian, J. Li, Z. Zhang, G. Sun, B. Zhao, H. Ma, R. Wu, Z. Wei, Y. Liu, L. Liao, Y. Ye, Y. Huang, X. Xu, X. Duan, W. Ji, X. Duan, Van der Waals epitaxial growth of air-stable CrSe$_{2}$ nanosheets with thickness-tunable magnetic order. \emph{Nat. Mater.} \textbf{20}, 818-825 (2021).
\bibitem{huang2017layer} B. Huang, G. Clark, E. Navarro-Moratalla, D. R. Klein, R. Cheng, K. L. Seyler, D. Zhong, E. Schmidgall, M. A. McGuire, D. H. Cobden, W. Yao, D. Xiao, P. Jarillo-Herrero, X. Xu, Layer-dependent ferromagnetism in a van der Waals crystal down to the monolayer limit. \emph{Nature} \textbf{546}, 270-273 (2017).
\bibitem{huang2018electrical} B. Huang, G. Clark, D. R. Klein, D. MacNeill, E. Navarro-Moratalla, K. L. Seyler, N. Wilson, M. A. McGuire, D. H. Cobden, D. Xiao, W. Yao, P. Jarillo-Herrero, X. Xu, Electrical control of 2D magnetism in bilayer CrI$_{3}$. \emph{Nat. Nanotechnol.} \textbf{13}, 544-548 (2018).
\bibitem{jiang2018controlling} S. Jiang, L. Li, Z. Wang, K. F. Mak, J. Shan, Controlling magnetism in 2D CrI$_{3}$ by electrostatic doping. \emph{Nat. Nanotechnol.} \textbf{13}, 549-553 (2018).
\bibitem{jiang2018electric} S. Jiang, J. Shan, K. F. Mak, Electric-field switching of two-dimensional van der Waals magnets. \emph{Nat. Mater.} \textbf{17}, 406-410 (2018).
\bibitem{webster2018strain} L. Webster, J.-A. Yan, Strain-tunable magnetic anisotropy in monolayer CrCl$_{3}$, CrBr$_{3}$, and CrI$_{3}$. \emph{Phys. Rev. B.} \textbf{98}, 144411 (2018).
\bibitem{wu2019strain} Z. Wu, J. Yu, S. Yuan, Strain-tunable magnetic and electronic properties of monolayer CrI$_{3}$. \emph{Phys. Chem. Chem. Phys.} \textbf{21}, 7750-7755 (2019).
\bibitem{song2019switching} T. Song, Z. Fei, M. Yankowitz, Z. Lin, Q. Jiang, K. Hwangbo, Q. Zhang, B. Sun, T. Taniguchi, K. Watanabe, M. A. McGuire, D. Graf, T. Cao, J.-H. Chu, D. H. Cobden, C. R. Dean, D. Xiao, X. Xu, Switching 2D magnetic states via pressure tuning of layer stacking. \emph{Nat. Mater.} \textbf{18}, 1298-1302 (2019).
\bibitem{cenker2022reversible} J. Cenker, S. Sivakumar, K. Xie, A. Miller, P. Thijssen, Z. Liu, A. Dismukes, J. Fonseca, E. Anderson, X. Zhu, X. Roy, D. Xiao, J.-H. Chu, T. Cao, X. Xu, Reversible strain-induced magnetic phase transition in a van der Waals magnet. \emph{Nat. Nanotechnol.} \textbf{17}, 256-261 (2022).
\bibitem{wilson2021interlayer} N. P. Wilson, K. Lee, J. Cenker, K. Xie, A. H. Dismukes, E. J. Telford, J. Fonseca, S. Sivakumar, C. Dean, T. Cao, X. Roy, X. Xu, X. Zhu, Interlayer electronic coupling on demand in a 2D magnetic semiconductor. \emph{Nat. Mater.} \textbf{20}, 1657-1662 (2021).
\bibitem{lee2013crystal} D. S. Lee, T.-H. Kim, C.-H. Park, C.-Y. Chung, Y. S. Lim, W.-S. Seo, H.-H. Park, Crystal structure, properties and nanostructuring of a new layered chalcogenide semiconductor, Bi$_{2}$MnTe$_{4}$. \emph{CrystEngComm} \textbf{15}, 5532-5538 (2013).
\bibitem{otrokov2019prediction} M. M. Otrokov, I. I. Klimovskikh, H. Bentmann, D. Estyunin, A. Zeugner, Z. S. Aliev, S. Ga$\ss$, A. U. B. Wolter, A. V. Koroleva, A. M. Shikin, M. Blanco-Rey, M. Hoffmann, I. P. Rusinov, A. Yu. Vyazovskaya, S. V. Eremeev, Yu. M. Koroteev, V. M. Kuznetsov, F. Freyse, J. S$\acute{a}$nchez-Barriga, I. R. Amiraslanov, M. B. Babanly, N. T. Mamedov, N. A. Abdullayev, V. N. Zverev, A. Alfonsov, V. Kataev, B. B$\ddot{u}$chner, E. F. Schwier, S. Kumar, A. Kimura, L. Petaccia, G. Di Santo, R. C. Vidal, S. Schatz, K. Ki$\ss$ner, M. $\ddot{U}$nzelmann, C. H. Min, S. Moser, T. R. F. Peixoto, F. Reinert, A. Ernst, P. M. Echenique, A. Isaeva, E. V. Chulkov, Prediction and observation of an antiferromagnetic topological insulator. \emph{Nature} \textbf{576}, 416-422 (2019).
\bibitem{zhang2019topological} D. Zhang, M. Shi, T. Zhu, D. Xing, H. Zhang, J. Wang, Topological axion states in the magnetic insulator MnBi$_{2}$Te$_{4}$ with the quantized magnetoelectric effect. \emph{Phys. Rev. Lett.} \textbf{122}, 206401 (2019).
\bibitem{li2019intrinsic} J. Li, Y. Li, S. Du, Z. Wang, B.-L. Gu, S.-C. Zhang, K. He, W. Duan, Y. Xu, Intrinsic magnetic topological insulators in van der Waals layered MnBi$_{2}$Te$_{4}$-family materials. \emph{Sci. Adv.} \textbf{5}, eaaw5685 (2019).
\bibitem{aliev2019novel} Z. S. Aliev, I. R. Amiraslanov, D. I. Nasonova, A. V. Shevelkov, N. A. Abdullayev, Z. A. Jahangirli, E. N. Orujlu, M. M. Otrokov, N. T. Mamedov, M. B. Babanly, E. V. Chulkov, Novel ternary layered manganese bismuth tellurides of the MnTe-Bi$_{2}$Te$_{3}$ system: Synthesis and crystal structure. \emph{J. Alloys Compd.} \textbf{789}, 443-450 (2019).
\bibitem{otrokov2019unique} M. M. Otrokov, I. P. Rusinov, M. Blanco-Rey, M. Hoffmann, A. Y. Vyazovskaya, S. V. Eremeev, A. Ernst, P. M. Echenique, A. Arnau, E. V. Chulkov, Unique thickness-dependent properties of the van der Waals interlayer antiferromagnet MnBi$_{2}$Te$_{4}$ films. \emph{Phys. Rev. Lett.} \textbf{122}, 107202 (2019).
\bibitem{147} C. Hu, K. N. Gordon, P. Liu, J. Liu, X. Zhou, P. Hao, D. Narayan, E. Emmanouilidou, H. Sun, Y. Liu, H. Brawer, A. P. Ramirez, L. Ding, H. Cao, Q. Liu, D. Dessau, N. Ni, A van der Waals antiferromagnetic topological insulator with weak interlayer magnetic coupling. \emph{Nat. Commun.} \textbf{11}, 97 (2020).
\bibitem{wu2019natural} J. Wu, F. Liu, M. Sasase, K. Ienaga, Y. Obata, R. Yukawa, K. Horiba, H. Kumigashira, S. Okuma, T. Inoshita, H. Hosono, Natural van der Waals heterostructural single crystals with both magnetic and topological properties. \emph{Sci. Adv.} \textbf{5}, eaax9989 (2019).
\bibitem{1813} C. Hu, L. Ding, K. N. Gordon, B. Ghosh, H.-J. Tien, H. Li, A. G. Linn, S.-W. Lien, C.-Y. Huang, S. Mackey, J. Liu, P. V. S. Reddy, B. Singh, A. Agarwal, A. Bansil, M. Song, D. Li, S.-Y. Xu, H. Lin, H. Cao, T.-R. Chang, D. Dessau, N. Ni, Realization of an intrinsic ferromagnetic topological state in MnBi$_{8}$Te$_{13}$. \emph{Sci. Adv.} \textbf{6}, eaba4275 (2020).
\bibitem{klimovskikh2020tunable} I. I. Klimovskikh, M. M. Otrokov, D. Estyunin, S. V. Eremeev, S. O. Filnov, A. Koroleva, E. Shevchenko, V. Voroshnin, A. G. Rybkin, I. P. Rusinov, M. Blanco-Rey, M. Hoffmann, Z. S. Aliev, M. B. Babanly, I. R. Amiraslanov, N. A. Abdullayev, V. N. Zverev, A. Kimura, O. E. Tereshchenko, K. A. Kokh, L. Petaccia, G. Di Santo, A. Ernst, P. M. Echenique, N. T. Mamedov, A. M. Shikin, E. V. Chulkov, Tunable 3D/2D magnetism in the (MnBi$_{2}$Te$_{4}$)(Bi$_{2}$Te$_{3}$)$_{m}$ topological insulators family. \emph{npj Quantum Mater.} \textbf{5}, 54 (2020).
\bibitem{ding2020crystal} L. Ding, C. Hu, F. Ye, E. Feng, N. Ni, H. Cao, Crystal and magnetic structures of magnetic topological insulators  MnBi$_{2}$Te$_{4}$ and MnBi$_{4}$Te$_{7}$. \emph{Phys. Rev. B} \textbf{101}, 020412 (2020).
\bibitem{shi2019magnetic} M. Shi, B. Lei, C. Zhu, D. Ma, J. Cui, Z. Sun, J. Ying, X. Chen, Magnetic and transport properties in the magnetic topological insulators MnBi$_{2}$Te$_{4}$(Bi$_{2}$Te$_{3}$)$_{n}$ (n= 1, 2). \emph{Phys. Rev. B} \textbf{100}, 155144 (2019).
\bibitem{chen2019topological} Y. J. Chen, L. X. Xu, J. H. Li, Y. W. Li, H. Y. Wang, C. F. Zhang, H. Li, Y. Wu, A. J. Liang, C. Chen, S. W. Jung, C. Cacho, Y. H. Mao, S. Liu, M. X. Wang, Y. F. Guo, Y. Xu, Z. K. Liu, L. X. Yang, Y. L. Chen, Topological electronic structure and its temperature evolution in antiferromagnetic topological insulator MnBi$_{2}$Te$_{4}$. \emph{Phys. Rev. X} \textbf{9}, 041040 (2019).
\bibitem{lee2019spin} S. H. Lee, Y. Zhu, Y. Wang, L. Miao, T. Pillsbury, H. Yi, S. Kempinger, J. Hu, C. A. Heikes, P. Quarterman, W. Ratcliff, J. A. Borchers, H. Zhang, X. Ke, D. Graf, N. Alem, C.-Z. Chang, N. Samarth, Z. Mao, Spin scattering and noncollinear spin structure-induced intrinsic anomalous Hall effect in antiferromagnetic topological insulator MnBi$_{2}$Te$_{4}$. \emph{Phys. Rev. Res.} \textbf{1}, 012011 (2019).
\bibitem{tian2019magnetic} S. Tian, S. Gao, S. Nie, Y. Qian, C. Gong, Y. Fu, H. Li, W. Fan, P. Zhang, T. Kondo, S. Shin, J. Adell, H. Fedderwitz, H. Ding, Z. Wang, T. Qian, H. Lei, Magnetic topological insulator MnBi$_{6}$Te$_{10}$ with a zero-field ferromagnetic state and gapped Dirac surface states. \emph{Phys. Rev. B} \textbf{102}, 035144 (2020).
\bibitem{gordon2019strongly} K. N. Gordon, H. Sun, C. Hu, A. G. Linn, H. Li, Y. Liu, P. Liu, S. Mackey, Q. Liu, N. Ni, D. Dessau, Strongly gapped topological surface states on protected surfaces of antiferromagnetic MnBi$_4$Te$_7$ and MnBi$_6$Te$_{10}$. arXiv:1910.13943 [cond-mat.str-el] (30 October 2019).
\bibitem{deng2020quantum} Y. Deng, Y. Yu, M. Z. Shi, Z. Guo, Z. Xu, J. Wang, X. H. Chen, Y. Zhang, Quantum anomalous Hall effect in intrinsic magnetic topological insulator MnBi$_2$Te$_4$. \emph{Science} \textbf{367}, 895-900 (2020).
\bibitem{liu2020robust} C. Liu, Y. Wang, H. Li, Y. Wu, Y. Li, J. Li, K. He, Y. Xu, J. Zhang, Y. Wang, Robust axion insulator and Chern insulator phases in a two-dimensional antiferromagnetic topological insulator. \emph{Nat. Mater.} \textbf{19}, 522-527 (2020).
\bibitem{ge2020high} J. Ge, Y. Liu, J. Li, H. Li, T. Luo, Y. Wu, Y. Xu, J. Wang, High-Chern-number and high-temperature quantum Hall effect without Landau levels. \emph{Natl. Sci. Rev.} \textbf{7}, 1280-1287 (2020).
\bibitem{gao2021layer} 1. A. Gao, Y.-F. Liu, C. Hu, J.-X. Qiu, C. Tzschaschel, B. Ghosh, S.-C. Ho, D. B$\acute{e}$rub$\acute{e}$, R. Chen, H. Sun, Z. Zhang, X.-Y. Zhang, Y.-X. Wang, N. Wang, Z. Huang, C. Felser, A. Agarwal, T. Ding, H.-J. Tien, A. Akey, J. Gardener, B. Singh, K. Watanabe, T. Taniguchi, K. S. Burch, D. C. Bell, B. B. Zhou, W. Gao, H.-Z. Lu, A. Bansil, H. Lin, T.-R. Chang, L. Fu, Q. Ma, N. Ni, S.-Y. Xu, Layer Hall effect in a 2D topological axion antiferromagnet. \emph{Nature} \textbf{595} 521-525 (2021).
\bibitem{yan2021delicate} C. Yan, Y. Zhu, S. Fernandez-Mulligan, E. Green, R. Mei, B. Yan, C. Liu, Z. Mao, S. Yang, Delicate Ferromagnetism in MnBi$_6$Te$_{10}$. arXiv:2107.08137 [cond-mat.mtrl-sci] (16 July 2021).
\bibitem{chen2019intrinsic} B. Chen, F. Fei, D. Zhang, B. Zhang, W. Liu, S. Zhang, P. Wang, B. Wei, Y. Zhang, Z. Zuo, J. Guo, Q. Liu, Z. Wang, X. Wu, J. Zong, X. Xie, W. Chen, Z. Sun, S. Wang, Y. Zhang, M. Zhang, X. Wang, F. Song, H. Zhang, D. Shen, B. Wang, Intrinsic magnetic topological insulator phases in the Sb doped MnBi$_2$Te$_4$ bulks and thin flakes. \emph{Nat. Commun.} \textbf{10}, 4469 (2019).
\bibitem{yan2019evolution} J.-Q. Yan, S. Okamoto, M. A. McGuire, A. F. May, R. J. McQueeney, B. C. Sales, Evolution of structural, magnetic, and transport properties in MnBi$_{2-x}$Sb$_{x}$Te$_4$. \emph{Phys. Rev. B} \textbf{100}, 104409 (2019).
\bibitem{liu2021site} Y. Liu, L.-L. Wang, Q. Zheng, Z. Huang, X. Wang, M. Chi, Y. Wu, B. C. Chakoumakos, M. A. McGuire, B. C. Sales, W. Wu, J. Yan, Site mixing for engineering magnetic topological insulators. \emph{Phys. Rev. X} \textbf{11}, 021033 (2021).
\bibitem{hu2021tuning} C. Hu, S.-W. Lien, E. Feng, S. Mackey, H.-J. Tien, I. I. Mazin, H. Cao, T.-R. Chang, N. Ni, Tuning magnetism and band topology through antisite defects in Sb-doped MnBi$_{4}$Te$_{7}$. \emph{Phys. Rev. B} \textbf{104}, 054422 (2021).
\bibitem{xie2021charge} H. Xie, F. Fei, F. Fang, B. Chen, J. Guo, Y. Du, W. Qi, Y. Pei, T. Wang, M. Naveed, others, Charge carrier mediation and ferromagnetism induced in MnBi$_{6}$Te$_{10}$ magnetic topological insulators by antimony doping. \emph{J. Phys. D: Appl. Phys} \textbf{55}, 104002 (2021).




\bibitem{zhang2021pressure} Z. Zhang, Z. Chen, Y. Zhou, Y. Yuan, S. Wang, J. Wang, H. Yang, C. An, L. Zhang, X. Zhu, Y. Zhou, X. Chen, J. Zhou, Z. Yang, Pressure-induced reemergence of superconductivity in the topological kagome metal CsV$_{3}$Sb$_{5}$. \emph{Phys. Rev. B} \textbf{103}, 224513 (2021).
\bibitem{chen2019suppression}  K. Chen, B. Wang, J.-Q. Yan, D. Parker, J.-S. Zhou, Y. Uwatoko, J.-G. Cheng, Suppression of the antiferromagnetic metallic state in the pressurized MnSb$_{4}$Te$_{7}$ single crystal. \emph{Phys. Rev. Mater.} \textbf{3}, 094201 (2019).

\bibitem{pei2020pressure} C. Pei, Y. Xia, J. Wu, Y. Zhao, L. Gao, T. Ying, B. Gao, N. Li, W. Yang, D. Zhang, H. Gou, Y. Chen, H. Hosono, G. Li, Y. Qi, Pressure-Induced topological and structural Phase transitions in an antiferromagnetic topological insulator. \emph{Chin. Phys. Lett.} \textbf{37}, 066401 (2020).
\bibitem{pei2022pressure} C. Pei, M. Xi, Q. Wang, W. Shi, L. Gao, Y. Zhao, S. Tian, W. Cao, C. Li, M. Zhang, S. Zhu, Y. Chen, H. Lei, Y. Qi, Pressure-induced superconductivity and structural phase transitions in magnetic topological insulator candidate MnSb$_{4}$Te$_{7}$. arXiv:2201.07635 [cond-mat.supr-con] (19 January 2022).
\bibitem{shao2021pressure} J. Shao, Y. Liu, M. Zeng, J. Li, X. Wu, X.-M. Ma, F. Jin, R. Lu, Y. Sun, M. Gu, K. Wang, W. Wu, L. Wu, C. Liu, Q. Liu, Y. Zhao, Pressure-tuned intralayer exchange in superlattice-like MnBi$_{2}$Te$_{4}$/(Bi$_{2}$Te$_{3}$)$_n$ topological insulators. \emph{Nano Lett.} \textbf{21}, 5874-5880 (2021).
\bibitem{vinokurova1976pressure} L. Vinokurova, A. Vlasov, M. Pardavi-Horv$\acute{a}$th, Pressure effects on magnetic phase transitions in FeRh and FeRhIr alloys. \emph{Phys. Status Solidi B} \textbf{78}, 353-357 (1976).
\bibitem{stern2017giant} E. Stern-Taulats, T. Cast$\acute{a}$n, A. Planes, L. H. Lewis, R. Barua, S. Pramanick, S. Majumdar, L. Ma$\tilde{n}$osa, Giant multicaloric response of bulk Fe$_{49}$Rh$_{51}$. \emph{Phys. Rev. B} \textbf{95}, 104424 (2017).
\bibitem{andrei2007magnetic} A. A. Katanin, V. Y. Irkhin, Magnetic order and spin fluctuations in low-dimensional insulating systems. \emph{Phys.-Usp.} \textbf{50}, 613 (2007).
\bibitem{stoner1948mechanism} E. C. Stoner, E. Wohlfarth, A mechanism of magnetic hysteresis in heterogeneous alloys. \emph{Philos. Trans. Royal Soc. A} \textbf{240}, 599-42 (1948).
\bibitem{yokogawa2007solidification} K. Yokogawa, K. Murata, H. Yoshino, S. Aoyama, Solidification of high-pressure medium Daphne 7373. \emph{Jpn. J. Appl. Phys.} \textbf{46}, 3636 (2007).
\bibitem{clark1978pressure} M. Clark, T. Smith, Pressure dependence of T$_c$ for lead. \emph{J. Low Temp. Phys.} \textbf{32}, 495-503 (1978).

\end{thebibliography}

\end{document}